\documentclass{aa}
\usepackage{graphicx}

\begin{document}

\title{Nonlinear magnetic diffusion and magnetic helicity transport in
galactic dynamos}
\author{N.\, Kleeorin \inst{1} \and D.\, Moss \inst{2} \and I.\,
Rogachevskii\inst{1}  \and D.\, Sokoloff\inst{3}} \offprints{I.
Rogachevskii} \institute{Department of Mechanical Engineering,
Ben-Gurion
University of Negev, POB 653,  84105 Beer-Sheva, Israel\\
\email{nat@menix.bgu.ac.il; gary@menix.bgu.ac.il} \and Department
of Mathematics, University of Manchester, Manchester M13
9PL, UK \\
\email{moss@maths.man.ac.uk} \and Department of Physics, Moscow
State University, Moscow
119992, Russia \\
\email{sokoloff@dds.srcc.msu.su}}

\date{Received ; accepted}

\abstract{We have extended our previous mean-field galactic dynamo
model which included algebraic and dynamic alpha nonlinearities
(Kleeorin et al. 2002), to include also a quenching of turbulent
diffusivity. We readily obtain equilibrium states for the
large-scale magnetic field in the local disc dynamo model, and
these fields have strengths that are comparable to the
equipartition field strength. We find that the algebraic
nonlinearity alone (i.e. quenching of both the $ \alpha $ effect
and turbulent magnetic diffusion) cannot saturate the growth of
the mean magnetic field; only the combined effect of algebraic and
dynamic nonlinearities can limit the growth of the mean magnetic
field. However, in contrast to our earlier work without quenching
of the turbulent diffusivity, we cannot now find satisfactory
solutions in the no-$z$ approximation to the axisymmetric galactic dynamo
problem.

}

\maketitle

\section{Introduction}

Spiral galaxies possess large-scale magnetic fields whose spatial
scale is comparable with galactic radii (see for review Beck et
al., 1996). Galactic magnetic fields mainly lie in the galactic
plane and the corresponding magnetic lines are usually roughly of
spiral form. This form can be substantially distorted in the
presence of strong noncircular motions, e.g. in barred galaxies,
see Beck et al. (1999, 2002), Moss et al.(2001).

Galactic magnetic fields are believed to originate in a galactic
dynamo, driven by the joint action of the mean hydrodynamic
helicity of interstellar turbulence and differential rotation. The
linear stage of galactic dynamo action seems now to be
well-understood, see, e.g., Ruzmaikin et al. (1988). The
conventional approach to the nonlinear stage of the galactic
dynamo is based on comparison of the relative intensity of three
quantities participating in dynamo action, namely the differential
rotation $\delta \Omega$, turbulent diffusivity $\eta_T$ and
$\alpha$-effect, each of which can be associated with a typical
velocity: $200$ km s$^{-1}$, $10$ km s$^{-1}$ and $1$ km s$^{-1}$
respectively. Because the typical velocity associated with
$\alpha$ is the smallest, the mean hydrodynamic helicity is
believed to be the weakest part of the self-excitation chain, and
a scenario of nonlinear dynamo limitation via $\alpha$-quenching
is the most commonly adopted.

A simple version of $\alpha$-quenching prescribes the mean
hydrodynamic helicity to be a decreasing function of mean magnetic
field strength $\bf B$. The critical magnetic field strength
$B_{\rm cr}$, at which quenching becomes significant, is estimated
conventionally from equipartition with the kinetic energy of
interstellar turbulence, $B^2_{\rm eq} \sim 4\pi\rho v_t^2$. When
applied to specific galaxies, this picture results in robust
magnetic field models which are compatible with observations.
However the picture is obviously oversimplified and various
attempts to suggest a more developed version of nonlinear galactic
dynamo theory have been undertaken by several authors. In
particular, Vainstein and Cattaneo (1992) and Gruzinov and Diamond
(1995) have claimed that in fact $B_{\rm cr}$ is much lower then
the equipartition value, e.g. $B_{\rm cr} = B_{\rm eq} Rm^{-1/2}$,
where $Rm$ is the magnetic Reynolds number. In galaxies, magnetic
Reynolds numbers are very large, $Rm \approx 10^6$ even if the
ambipolar diffusivity coefficient is used, so it was claimed that
dynamo action saturates at a magnetic field strength that is much
lower than both the equipartition value, and also the
large-scale field strengths observed in nearby spiral galaxies.
This result follows from investigations that accept the idea of
magnetic helicity conservation. The galactic dynamo produces a
large-scale magnetic field with nonvanishing magnetic helicity,
and  when considering detailed magnetic helicity conservation in a
given volume, the above `catastrophic' estimate for $B_{\rm cr}$
results.

The evolution of magnetic helicity appears however to be a more
complicated process than can simply be described by a balance of
magnetic helicity in a given volume, and it is necessary to take
into account, as for the evolution of other conserved quantities,
transport by the fluid flows including turbulent transport of
magnetic helicity through the galactic boundaries and the
destruction of {\it mean} magnetic helicity by turbulent
diffusion. The governing equation for magnetic helicity was
suggested by Kleeorin \& Ruzmaikin (1982; see also the discussion
by Zeldovich et al. 1983) for an isotropic turbulence, and
investigated by Kleeorin et al. (1995) for stellar dynamos, and
self-consistently derived by Kleeorin \& Rogachevskii (1999) for
an arbitrary anisotropic turbulence.  A quantitative model for the
flux of magnetic helicity was proposed by Kleeorin \& Rogachevskii
(1999) and Kleeorin et al. (2000). Note that Schmalz \& Stix
(1991), Covas et al. (1998) and Blackman \& Brandenburg (2002)
have also investigated related solar dynamo models that included a dynamical
equation describing the evolution of magnetic helicity. Magnetic helicity transport
through the boundary of a dynamo region is reported by Chae (2001)
to be observable at the solar surface. The role of a flux of
magnetic helicity in the dynamics of the mean magnetic field in
accretion discs was also discussed by Vishniac \& Cho (2001).

The equation governing the magnetic helicity is much more complex than the
conventional parametrization used to represent $\alpha$-quenching.
Kleeorin et al. (2000) suggested that a nonlinear galactic dynamo
governed by a consistently derived equation for magnetic helicity
results in a steady magnetic field comparable with the
equipartition magnetic field estimate, and Kleeorin et al. (2002)
demonstrated that a detailed galactic dynamo model based on the
equation under discussion gives results very similar to one based
on conventional $\alpha$-quenching. In other words, the real
physical description of the nonlinear stage of galactic dynamo is
quite complicated but, if we are interested in pragmatic results
only, an adequate description can be found by using only a
conventional $\alpha$-quenching.

We stress that the scenario of Kleeorin et al. (2000, 2002) does
not include all possible types of nonlinear processes which can
occur at the nonlinear dynamo stage (see, e.g., Brandenburg \&
Subramanian 2000; Brandenburg \& Dobler 2001), but rather is
restricted by a minimal number of processes involved in the
magnetic helicity conservation. In particular, we consider
transport of mean helicity of small-scale magnetic field only and
note that this helicity may be transported out of the galactic
disc without significant losses of large-scale magnetic field.
Moreover, this picture is formulated in terms of mean-field
electrodynamics, which gives a natural constraint on the
description of small-scale details. When attention is focussed on
these details (e.g. Brandenburg \& Sokoloff 2002; Blackman \&
Brandenburg 2002), the mean-field description should be considered
as a parametrization of the turbulence.

In the spirit of the basic ideas about the nonlinear saturation of
galactic dynamos, the analysis presented by Kleeorin et al. (2000,
2002) was restricted to the evolution of $\alpha$ only, while a
detailed simulation (e.g. Brandenburg \& Sokoloff 2002)
also demonstrates a quenching of the turbulent magnetic diffusivity.
A quantitative model for a nonlinear quenching of turbulent
magnetic diffusivity has been recently suggested by Rogachevskii
\& Kleeorin (2001).

The aim of the present paper is to include a turbulent magnetic
diffusivity quenching into the mean-field dynamo equations. This
effect is expected to be quite modest. Speaking pragmatically, we
do not know the turbulent magnetic diffusivity of interstellar
turbulence well enough to recognize its saturation, by some dozens
of percent. However, our analysis below demonstrates that the
problem is not restricted by some specific variation of magnetic
diffusivity coefficient. Because of nonlinear effects, the
turbulent magnetic diffusion coefficients for the two basic
magnetic field components, i.e. poloidal and toroidal, become
different (see Rogachevskii \& Kleeorin 2001). Since spatial
magnetic field distribution is nonuniform, a nonuniform magnetic
diffusivity saturation arises, that results in new terms in the
dynamo equations governing the nonlinear magnetic field
evolution. In general, the situation appears to be less trivial
then might be thought initially, and a quantitative
analysis of a specific dynamo model becomes desirable. Below we
present results of the corresponding analysis and numerical
simulations.

\section{The mean-field equations}

The mean-field dynamo equation (e.g. Krause \& R\"adler 1980) is
\begin{eqnarray}
\frac{\partial {\vec{B}}}{\partial t}= \vec{\nabla} {\bf \times}
(\vec{V} {\bf \times} \vec{B} + \vec{\cal E} - \eta \,
\vec{\nabla} {\bf \times} \vec{B}) \;, \label{E1}
\end{eqnarray}
where $ {\bf V} $ is a mean velocity ({\em e.g.,} the differential
rotation), $ \eta $ is the magnetic diffusion due to the
electrical conductivity of fluid together with ambipolar
diffusion, $ \vec{\cal E} = \langle {\bf u} \times {\bf b} \rangle
$ is the mean electromotive force, $ {\bf u} $ and $ {\bf b} $ are
fluctuations of the velocity and magnetic field, respectively,
angular brackets denote averaging over an ensemble of
fluctuations. When a small-scale background turbulence (i.e.
turbulence with a zero mean magnetic field) is isotropic and the
energy of small-scale magnetic fluctuations of the background
turbulence is much smaller than that of the kinetic energy, the
mean electromotive force is given by
\begin{eqnarray}
{\cal E}_{i} = \alpha_{ij}(\vec{B}) B_{j} + (\vec{V}^{\rm
eff}(\vec{B}) {\bf \times} \vec{B})_{i} - \eta_{ij}(\vec{B})
(\vec{\nabla} {\bf \times} \vec{B})_{j} \;,
\label{E2}
\end{eqnarray}
where
\begin{eqnarray}
\eta_{ij}(\vec{B}) &=& [\eta_A(\vec{B}) - \eta_\beta (\vec{B})] \,
\delta_{ij} + \eta_\beta(\vec{B}) \, B_{ij} \;,
\label{E3}\\
\vec{V}^{\rm eff}(\vec{B}) &=& {1 \over 2} \eta_V(\vec{B})
{\vec{\nabla} B^2 \over B^2} \;, \label{E4}
\end{eqnarray}
(see Rogachevskii \& Kleeorin, 2001), $ B_{ij} = B_{i} B_{j} / B^2
,$ $ \, \alpha_{ij}(\vec{B}) = \alpha(\vec{B}) \delta_{ij} ,$ $ \,
\eta_V(\vec{B}) = \eta_A(\vec{B}) - \eta_B(\vec{B}) -
\eta_\beta(\vec{B}) $ and $\alpha(\vec{B})$, $\, \eta_A(\vec{B})
,$ $\,\eta_B(\vec{B})$, $ \, \eta_\beta(\vec{B}) $  are determined
by Eqs.~(\ref{A3}), (\ref{D1}), (\ref{D2}) and (\ref{D3D}),
respectively.

\subsection{The local thin-disc dynamo problem}
\label{basic}

We begin by considering the simplest local disc dynamo problem.
Using cylindrical polar coordinates $r, \phi, z$, from
Eqs.~(\ref{E1})-(\ref{E4}) we obtain the equations for the mean
radial field $B_r=R_\alpha b_r$ and toroidal field $B_\phi$ for
the local thin-disc axisymmetric $\alpha \Omega$-dynamo problem as
\begin{eqnarray}
{{\partial b_r} \over {\partial t}} &=& - (\alpha (\vec{B})
B_\phi)' + (\eta_{A}(\vec{B}) b_r')' - (V_{A}(\vec{B}) b_r)' \;,
\label{A1} \\
{{\partial B_\phi} \over {\partial t}} &=& D b_r +
(\eta_{B}(\vec{B}) B'_\phi)' \; \label{A2}
\end{eqnarray}
(Ruzmaikin et al. 1988, Rogachevskii \& Kleeorin 2001). Here a prime
denotes $
\partial/\partial z,$ $\alpha (\vec{B})$ is the total nonlinear $\alpha$
effect, $ \eta_{A}(\vec{B}) $ and $ \eta_{B}(\vec{B}) $ are the
nonlinear turbulent magnetic diffusion coefficients of poloidal
and toroidal mean magnetic fields, and the nonlinear function $
V_{A}(\vec{B}) \equiv [\eta_{A}(\vec{B}) - \eta_{B}(\vec{B})] (\ln
B)' $, with $B=|\vec{B}|.$ These nonlinearities are specified and
discussed in the next Section.

We adopt here the standard dimensionless form of the galactic
dynamo equations from Ruzmaikin et al. (1988); in particular,
length is measured in units of the disc thickness $h$, time in
units of $ h^{2} / \eta_{T} $ and $B$ is measured in units of the
equipartition energy $B_{\rm eq} = \sqrt{4 \pi \rho} \, u_\ast $,
$ \alpha $  is measured  in units of $ \alpha_\ast $ (the maximum
value of the hydrodynamic part of the $ \alpha $ effect), the
nonlinear turbulent magnetic diffusion coefficients $
\eta_{A}(\vec{B}) $ and $ \eta_{B}(\vec{B}) $ are measured in
units of $ \eta_{T} $. We define $R_\alpha = h \alpha_\ast /
\eta_{T} ,$ $\, R_\omega =r \, (d\Omega/dr) \, h^2/\eta_T$, and
the dynamo number $D=R_\omega R_\alpha $, where $l$ is the maximum
scale of the turbulent motions, $ Rm = l u_{\ast} / \eta $ is the
magnetic Reynolds number. Also $u_{\ast}$ is the characteristic
turbulent velocity at the scale $l$, $\rho$ is the gas density,
and the characteristic value of the turbulent magnetic diffusivity
$\eta_T=l u_{\ast}/3$.

Throughout this paper, unless otherwise stated, we measure
magnetic field in units of the equipartition value. Here we
assumed that the background turbulence (i.e., the turbulence with
a zero mean magnetic field) is isotropic and has only velocity
fluctuations, even though  a nonzero mean magnetic field can be
expected to produce an anisotropy of turbulence from the generated
magnetic fluctuations.

\section{The nonlinearities}
\label{nonlin}
\subsection{The nonlinear $\alpha$ effect}

The total nonlinear $\alpha$ effect is given by
\begin{eqnarray}
\alpha(\vec{B}) = \alpha^v + \alpha^m \;,
\label{AA3}
\end{eqnarray}
where $\alpha^v$ is the hydrodynamic part of the $\alpha$ effect,
and $\alpha^m$ is the magnetic part of the $\alpha$ effect. These
quantities are determined by the corresponding helicities and
quenching functions, $\phi_{v}(B)$ and $\phi_{m}(B) .$ In
particular, $\alpha^v = \chi^v \phi_{v}(B) ,$ $\alpha^m =
\chi^c(\vec{B}) \phi_{m}(B) $ and $\chi^v = - (\tau /3)
\langle~\vec{u}~\cdot~(\vec{\nabla} {\bf \times} \vec{u})~\rangle
,$ $ \; \chi^{c} \equiv (\tau / 12 \pi \rho) \langle \vec{b} \cdot
(\vec{\nabla} {\bf \times} \vec{b}) \rangle $ is related with
current helicity, where $ \tau $ is the correlation time of
turbulent velocity field and $\langle \vec{u} \cdot (\vec{\nabla}
{\bf \times} \vec{u}) \rangle$ is the hydrodynamic helicity. Thus,
\begin{eqnarray}
\alpha(\vec{B}) = \chi^v \phi_{v}(B) + \chi^c(\vec{B}) \phi_{m}(B)
\; , \label{A3}
\end{eqnarray}
where the quenching functions $\phi_{v}(B)$ and $\phi_{m}(B)$ are
given by
\begin{eqnarray}
\phi_{v}(B) &=& (1/7) [4 \phi_{m} (B) + 3 L(\sqrt{8} B)] \;,
\label{A4} \\
\phi_{m} (B) &=& (3 / \beta^{2}) (1 -\arctan (\beta) / \beta) \;
\label{A5}
\end{eqnarray}
(see Rogachevskii \& Kleeorin, 2000), where $\beta=\sqrt{8} B $
and $ L(\beta) = 1 - 2 \beta^{2} + 2 \beta^{4} \ln (1 +
\beta^{-2}) .$  Thus $\phi_{v}(B) = 2/\beta^2$ and $\phi_{m}(B) =
3/\beta^2$ for $\beta \gg 1 ;$ and $\phi_{v}(B) = 1-(6/5)\beta^2$
and $\phi_{m}(B) = 1-(3/5)\beta^2$ for $\beta \ll 1 .$ The
function $\chi^c(\vec{B})$ entering the magnetic part of the
$\alpha$ effect is determined by the dynamical
equation~(\ref{A6A}). Here $ \chi^v $ and $ \chi^c$ are measured
in units of $ \alpha_\ast $.

The function $\phi_{v}(B)$ describes conventional quenching of the
$ \alpha $ effect. A simple form of such a quenching, $\phi_v =
1/(1 + B^{2}) ,$ was introduced long ago (see, e.g., Iroshnikov,
1970). The splitting of the total $\alpha$ effect into the
hydrodynamic ($\alpha^v $) and magnetic ($\alpha^m $) parts was
first suggested by Frisch et al. (1975). The magnetic part
$\alpha^m$ includes two types of nonlinearity: the algebraic
quenching described by the function $\phi_{m}(B)$ (see e.g. Field
et al. 1999; Rogachevskii \& Kleeorin 2000, 2001) and the dynamic
nonlinearity which is determined by Eq. (\ref{A6A}).

\subsection{Nonlinear turbulent magnetic diffusion  coefficients
of the toroidal and poloidal mean magnetic fields}

The nonlinear turbulent magnetic diffusion coefficients of poloidal
and toroidal mean magnetic fields $ \eta_{A}(B) $ and $
\eta_{B}(B) $, and the nonlinear function $ V_{A}(B) \equiv
[\eta_{A}(B) - \eta_{B}(B)] (\ln B)' $ are given in dimensionless form
by
\begin{eqnarray}
\eta_{A}(B) &=& A_{1}(\sqrt{2}\beta) + (1/2) A_{2}(\beta) \;,
\label{D1} \\
\eta_{B}(B) &=& A_{1}(\sqrt{2}\beta) - (1/2) A_{2}(\beta) +
A_{2}(\sqrt{2}\beta)
\nonumber \\
& & + (\beta /\sqrt{2}) \Psi(\sqrt{2}\beta) \;,
\label{D2}\\
V_{A}(B) &=& [A_{2}(\beta) - A_{2}(\sqrt{2}\beta)
\nonumber \\
& & - (\beta /\sqrt{2}) \Psi(\sqrt{2}\beta)](\ln B)' \;,
\label{D3}
\end{eqnarray}
(see Rogachevskii \& Kleeorin, 2001), where $ \Psi(x) = (d/dx)
[A_{1}(x) + (1/2) A_{2}(x)] ,$ and
\begin{eqnarray}
\eta_\beta(\vec{B})= (1/2) A_2(\beta) \; .
\label{D3D}
\end{eqnarray}

The functions $ A_{1}(\beta) $ and $ A_{2}(\beta) $ are given by
\begin{eqnarray*}
A_{1}(\beta) &=& {6 \over 5} \biggl[{\arctan \beta \over \beta}
\biggl(1 + {5 \over 7 \beta^{2}} \biggr) + {1 \over 14} L(\beta) -
{5 \over 7\beta^{2}} \biggr]  \;,
\\
A_{2}(\beta) &=& - {6 \over 5} \biggl[{\arctan \beta \over \beta}
\biggl(1 + {15 \over 7 \beta^{2}} \biggr) - {2 \over 7} L(\beta) -
{15 \over 7\beta^{2}} \biggr]  \; .
\end{eqnarray*}
When $ \beta \ll 1 $ these functions are given approximately by
\begin{eqnarray*}
A_{1}(\beta) &=& 1 - (2 / 5) \beta^{2}  \;, \quad A_{2}(\beta)
= - (4 / 5) \beta^{2} \;,
\end{eqnarray*}
and for $ \beta \gg 1 $ they are given by
\begin{eqnarray*}
A_{1}(\beta) &=& 3 \pi / 5 \beta - 4 / 5 \beta^{2} \;, \quad
A_{2}(\beta) = - 3 \pi / 5 \beta + 14 / 5 \beta^{2} \; .
\end{eqnarray*}

The asymptotic formulas for the functions $ \eta_{A}(B) ,$ $
\eta_{B}(B) $ and $ V_{A}(B) $ for $ \beta \ll 1 $ are
\begin{eqnarray*}
\eta_{A}(B) &=& 1 - (6/5) \beta^{2} \;, \quad \eta_{B}(B) = 1 -
(18/5) \beta^{2} \;,
\\
V_{A}(B) &=& (12/5) \beta^{2} (\ln B)'  \;,
\end{eqnarray*}
and for $ \beta \gg 1 $
\begin{eqnarray*}
\eta_{A}(B) &=& {3 \pi \over 10 \beta} (\sqrt{2} - 1) \;,
\\
\eta_{B}(B) &=& {3 \pi \over 20 \beta} (2 - {1 \over \sqrt{2}})
\;,
\\
V_{A}(B) &=& - {3 \pi \over 5 \beta} (1 - {5 \sqrt{2} \over 8})
(\ln B)' \; .
\end{eqnarray*}

The quenching of the $ \alpha $ effect and the turbulent magnetic
diffusion are caused by the direct and indirect modification of
the electromotive force by the mean magnetic field. The indirect
modification of the electromotive force is caused by the effect of
the mean magnetic field on the velocity fluctuations (described by the
tensors $ \langle u_i u_j \rangle $) and on the magnetic fluctuations
(determined by the tensor $ \langle b_i b_j \rangle ),$ while the
direct modification is due to the effect of the mean magnetic
field on the cross-helicity tensor $ \langle u_i b_j \rangle $
(see, e.g., Rogachevskii \& Kleeorin 2000, 2001).

\subsection{The dynamical equation for the function
$\chi^c(\vec{B})$}

The function $\chi^c(\vec{B})$ entering the magnetic part of the
$\alpha$ effect is determined by the dynamical equation
\begin{eqnarray}
{{\partial \chi^{c}} \over {\partial t}} &=& - 4 \biggl({h \over
l} \biggr)^2 [\vec{\cal E} {\bf \cdot} \vec{B} + \vec{\nabla}
\cdot \vec{F}] - \vec{\nabla} \cdot [\vec{V} \chi^{c}
\nonumber \\
& & - \kappa \vec{\nabla}\chi^c] - \chi^{c} / T \; , \label{A6A}
\end{eqnarray}
where $ \vec{F} = C \chi^v \phi_{v}(B) \vec{B}^2 \eta_{A}(\vec{B})
(\vec{\nabla} \rho) / \rho $ is the nonadvective flux of the
magnetic helicity which serves as an additional nonlinear source
in the equation for $ \chi^{c} $, $ \, \vec{V} \chi^{c}$ is the
advective flux of the magnetic helicity and $  - \kappa
\vec{\nabla}\chi^c$ is the diffusive flux of the magnetic helicity
(see Kleeorin \& Rogachevskii, 1999; Kleeorin et al. 2000; 2002),
$\, \vec{V} = \vec{e}_{\phi} \, \Omega \, r $ is the differential
rotation, and $ T = (1/3) (l/h)^{2} Rm $. Eq.~(\ref{A6A}) was
obtained using arguments based on the magnetic helicity
conservation law (see Kleeorin \& Rogachevskii, 1999). The
function $\chi^{c}$ is proportional to the magnetic helicity,
$\chi^{c} = \chi^{m} / (18 \pi \eta_{T} \rho)$ (see e.g. Kleeorin
\& Rogachevskii, 1999), where $\chi^{m} = \langle \vec{a} \cdot
\vec{b} \rangle$ is the magnetic helicity and $\vec{a}$ is the
vector potential of small-scale magnetic field. Here we assume
that the helical part of the vector potential $\vec{a}$ is a locally
isotropic and homogeneous random field, which is a natural gauge
condition used in our approach. Thus, Eq.~(\ref{A6A}) describes
the evolution of magnetic helicity, i.e. its production,
dissipation and transport.

The turbulent diffusion of the magnetic helicity $\kappa$ plays an
important role and can be interpreted as follows. The random flows
existing in the interstellar medium consist of a combination of
small-scale motions, which are affected by magnetic forces
resulting in a steady-state of the dynamo, and a microturbulence
which is supported by a strong random driver (supernovae
explosions) which can be considered as independent of the galactic
magnetic field. The large-scale magnetic field is smoothed over
both kinds of turbulent fluctuations, while the small-scale
magnetic field is smoothed over microturbulent fluctuations only.
It is the smoothing over the microturbulent fluctuations that
gives the coefficient $\kappa$.

For galaxies the relaxation term $\chi^{c} / T$ is very small and
can be dropped in spite of the fact the small yet finite magnetic
diffusion is required for the reconnection of magnetic field
lines. For example, we will show below that the magnetic Reynolds
number, $ Rm ,$ does not enter into the steady state solution of
Eq.~(\ref{A6A}) in the limit of very large  $ Rm $, because of the
effect of the magnetic helicity flux. In particular, keeping the
term $\chi^{c} / T$  we obtain from Eq.~(\ref{A6A}) that in a
steady state
\begin{eqnarray}
\alpha = {\alpha_v + \tau_\ast \, Rm \, \phi_m(\vec{B}) \,
[\vec{B} \cdot (\vec{\nabla} {\bf \times} \vec{B}) - {\rm div} \,
\vec{F}_t] \over 1 + \tau_\ast \, Rm \, \phi_m(\vec{B}) \, B^2}
\;, \label{AA6A}
\end{eqnarray}
where $ \tau_\ast = l / u_\ast $ and $\vec{F}_t=\vec{F}+
\vec{V}\chi^c-\kappa\nabla\chi^c$ is the total flux
of the magnetic helicity.  In the limit of very large $ Rm ,$
Eq.~(\ref{AA6A}) reads
\begin{eqnarray}
\alpha = {\vec{B} \cdot (\vec{\nabla} {\bf \times} \vec{B}) - {\rm
div} \, \vec{F}_t \over B^2} \; . \label{AAA6A}
\end{eqnarray}
This implies that in this limit the total $\alpha$ is independent
of the magnetic Reynolds number.

In the local approximation Eq.~(\ref{A6A}) reads:
\begin{eqnarray}
{\partial \chi^c \over \partial t} &=& 4 \biggl({h \over l}
\biggr)^2 [\eta_{A}(\vec{B}) (B_\phi b'_r - B'_\phi b_r) -
\alpha(\vec{B}) B^2
\nonumber \\
& & + C{{\partial} \over {\partial z}} (|\chi^v(z)| \phi_{v}(B)
B^2 \eta_{A}(\vec{B}))] + (\kappa (\chi^c)')' \; .
\label{A6}
\end{eqnarray}
Here we do not take into account any inhomogeneity of the turbulent
magnetic diffusion at ${\bf B}=0$. The turbulent magnetic
diffusion is inhomogeneous due to inhomogeneity of the mean
magnetic field ${\bf B}$.

The turbulent magnetic diffusion $ \kappa $ of the magnetic
helicity (and the function $ \chi^c )$ can depend on the mean
magnetic field. The nonlinear quenching of the turbulent magnetic
diffusion of the magnetic helicity is given by
$\phi_{\kappa}(B)$,
\begin{eqnarray}
\phi_{\kappa}(B) = {1 \over 2} \biggl[1 + A_{1}(\sqrt{2}\beta) +
{1 \over 2} A_{2}(\sqrt{2}\beta) \biggr] \; .
\label{D10}
\end{eqnarray}
For $ \beta \ll 1 $ we have approximately
\begin{eqnarray*}
\phi_{\kappa}(B) = 1 - {3 \over 5} \beta^2 \;,
\end{eqnarray*}
and for $ \beta \gg 1 $ we have
\begin{eqnarray*}
\phi_{\kappa}(B)  = {1 \over 2} \biggl( 1 + {3 \pi \over 10
\sqrt{2} \beta} \biggr) \; .
\end{eqnarray*}
The turbulent magnetic diffusion $ \kappa(B) $ of the magnetic
helicity is determined by the tensor $ \tau \langle u_i u_j
\rangle .$
Thus, Eqs (\ref{A3}), (\ref{D1})--(\ref{D3}) and (\ref{A6})
contain the main nonlinearities.

\section{Equilibrium states of the local dynamo model}

\subsection{Asymptotic expansions and an equilibrium solution}

We now present asymptotic expansions for a galactic dynamo model
determined by Eqs. (\ref{A1}), (\ref{A2}) and (\ref{A6}). For the
$\alpha \Omega$-dynamo $ B \approx B_\phi .$ This assumption is
justified if $|D| \gg R_\alpha ,$ i.e. $|R_\omega| \gg 1 .$ In a
steady-state for fields of even parity with respect to the disc
plane, Eqs. (\ref{A1}), (\ref{A2}) and (\ref{A6}) with $ \kappa =
0 $ gives
\begin{eqnarray}
[\eta_{B}(B) B']^{2} + 2 C D \phi_{v}(B) \eta_{A}(B) B^2
|\chi^v(z)| = 0 \; . \label{C1}
\end{eqnarray}
The solution of Eq. (\ref{C1}) for negative $ D $ is given by
\begin{eqnarray}
\int_{0}^{B} G(\tilde B) \,d \tilde B = \sqrt{2 C |D|}
\int_{|z|}^{1} \sqrt{|\chi^v(\tilde z)|} \,d \tilde z \;,
\label{C2}
\end{eqnarray}
where $ G(B) = \eta_{B}(B) /    [\phi_{v}(B) \eta_{A}(B)
B^2]^{1/2} .$ For an arbitrary  profile $|\chi^v(z)|$, negative
dynamo number $D$ and for  $ B \gg 1/ \sqrt{8},$ there is an
explicit steady solution of this equation with the boundary
conditions $ B_{\phi}(z=1) = 0 $ and $ B_{\phi}'(z=0) = 0 $,
\begin{eqnarray}
B(z) \approx (2/5) C |D| \biggl( \int_{|z|}^{1}
\sqrt{|\chi^v(\tilde z)|} \,d \tilde z \biggr)^{2} \;, \label{C3}
\end{eqnarray}
where $z$ is measured in the units of $h ,$ and we have used that for $
B \gg 1/ \sqrt{8} $ we have $ \eta_{A}(B) \sim 2 / 5 \beta
,$ $ \eta_{B}(B) \sim 3 / 5 \beta ,$ $ \phi_{v}(B) \sim 2 /
\beta^{2} .$ Here $ \beta = \sqrt{8} B .$ In a steady state $
b_r(z) = (\eta_{B}(B) B')'  / |D| .$ For the specific choice of
the profile $|\chi^v(z)| =\sin^{2}(\pi z/ 2)$ we obtain
\begin{eqnarray}
B_\phi &\approx & {4 \over 25} C \, |D| \, B_{\rm eq} \cos^2 \,
\biggl({{\pi z} \over 2} \biggr) \;,
\label{C6} \\
B_r &\approx & - {1 \over |R_\omega|} B_{\rm eq} \cos^{-2} \,
\biggl({{\pi z} \over 2} \biggr) \;,
\label{C7}
\end{eqnarray}
where we have now restored the dimensional factor $B_{\rm eq} .$ The
boundary conditions for $B_{r}$ are $B_{r}(z=1) = 0$ and $
B_{r}'(z=0) = 0 .$ Note, however, that our asymptotic analysis
performed for $ B \gg 1/ \sqrt{8} $  is not valid in the vicinity
of the point $z=1$ because $B(z=1) = 0$.

\subsection{Numerical solutions for the one-dimensional model}

We found solutions of Eqs.~(\ref{A1}), (\ref{A2}) and (\ref{A6})
by step by step integration, from arbitrarily chosen initial
conditions. Various properties of the solutions for the mean
magnetic field and the z-profiles of the main nonlinearities -- the
$\alpha$--effect, turbulent magnetic diffusion of the toroidal and
poloidal fields -- are illustrated in Figs.~1 -- 6. It can be seen
clearly that the field strength ($\sim |B_\phi|$) is typically of
order 1 (equipartition), and increases with $|D|$.

\begin{figure}
\centering
\includegraphics[width=8cm]{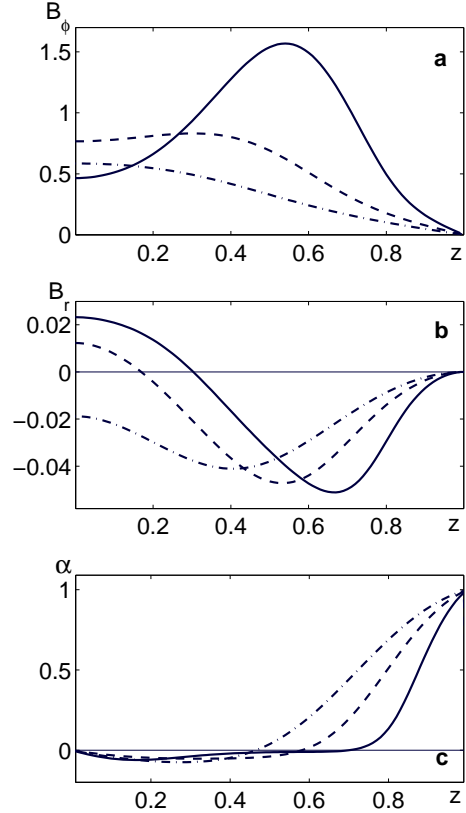}
\caption{\label{Fig1} The $z$-dependence of solutions for the
local model; $h/l=5$; $C=-0.1$, $\kappa=0.3$ and
$|\chi^v(z)|=\sin^2(\pi z/2)$. The various curves indicate results
with dynamo numbers $D=-20$ (dashed-dotted), $D=-40$ (dashed) and
$D=-100$ (solid). a) the toroidal magnetic field $B_{\phi}(z)$, b)
the poloidal magnetic field $B_{r}(z)$, c) the total
$\alpha$-effect, $\alpha(z)$.}
\end{figure}

\begin{figure}
\centering
\includegraphics[width=8cm]{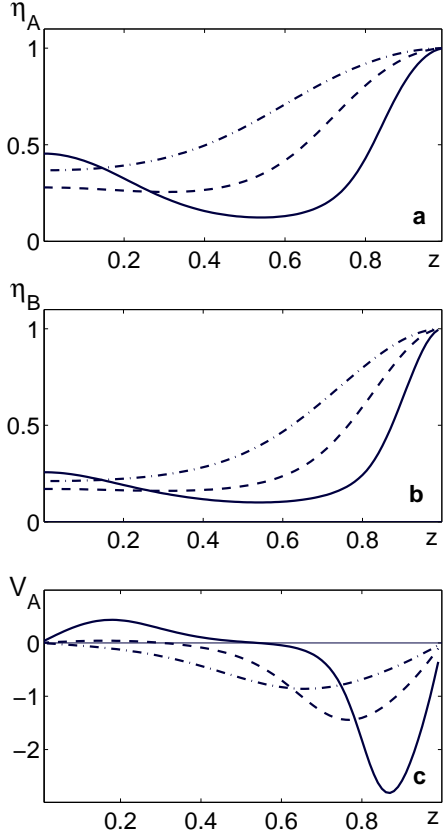}
\caption{\label{Fig2} As Fig.~1. The functions: a) $\eta_{A}(z)$,
b) $\eta_{B}(z)$, c) $V_{A}(z)$.}
\end{figure}

\begin{figure}
\centering
\includegraphics[width=8cm]{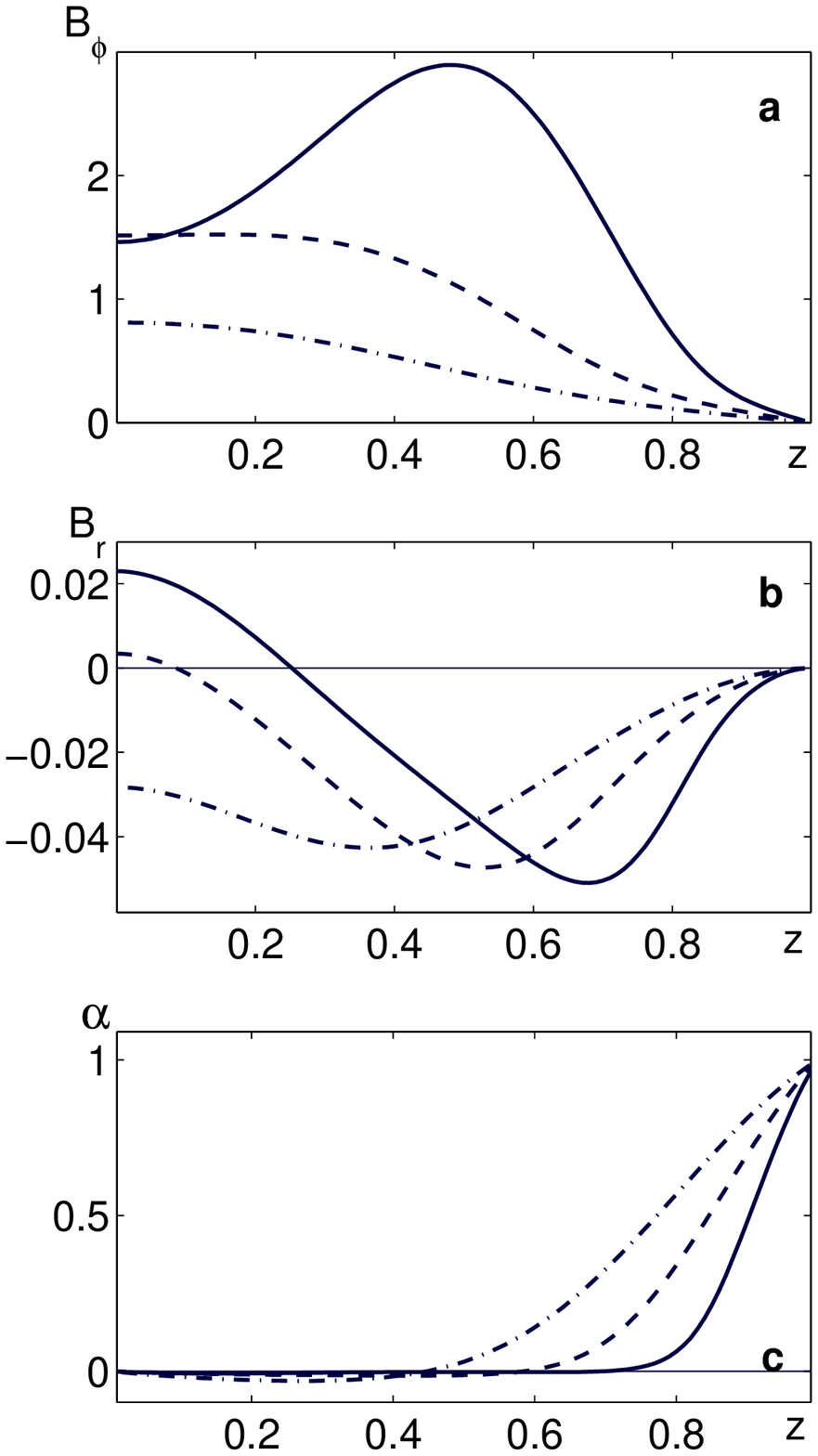}
\caption{\label{Fig3} As Fig.~1, with $C=0.1$.}
\end{figure}

\begin{figure}
\centering
\includegraphics[width=8cm]{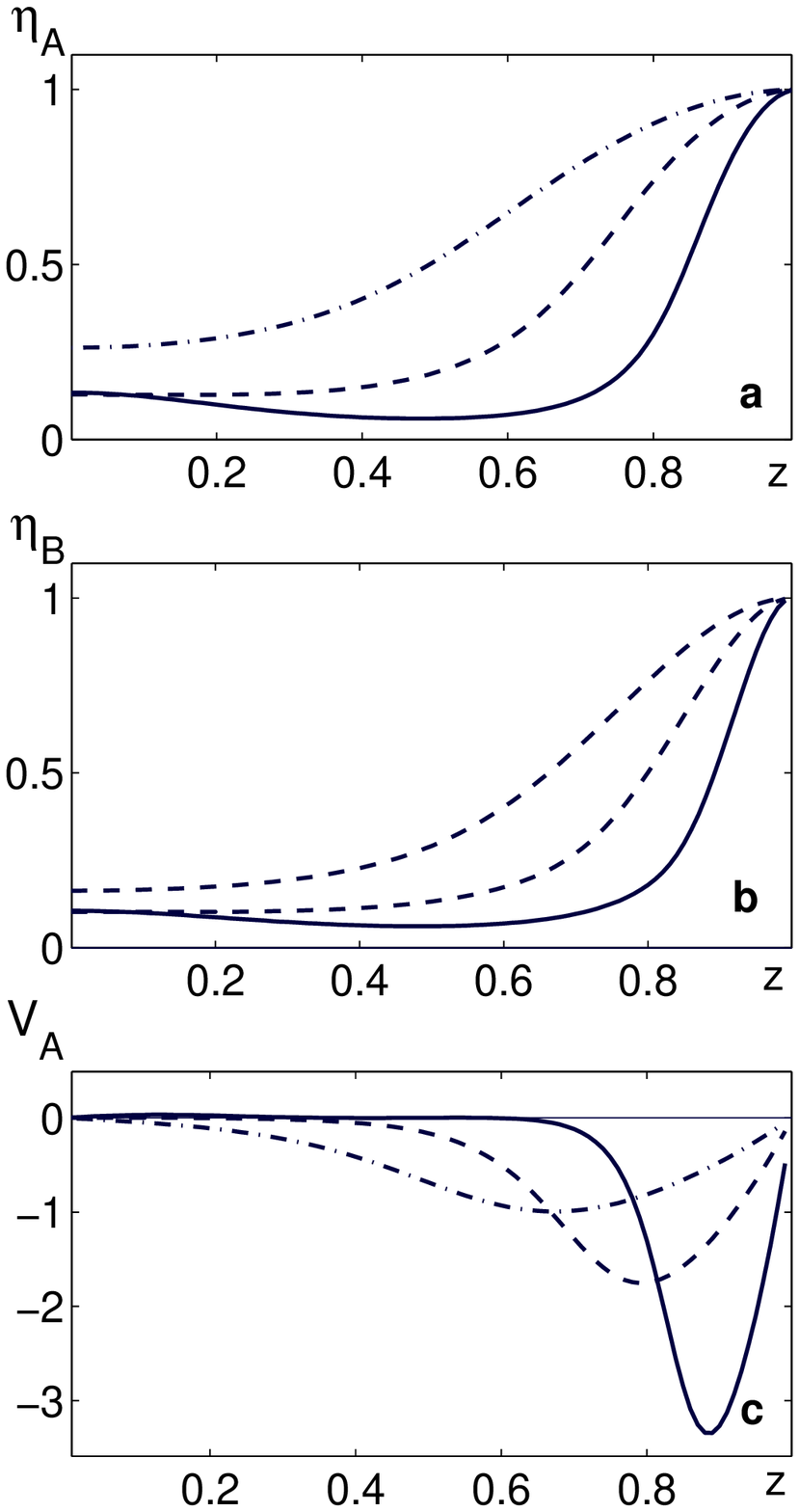}
\caption{\label{Fig4}  As Fig.~2, with $C=0.1$.}
\end{figure}

\begin{figure}
\centering
\includegraphics[width=8cm]{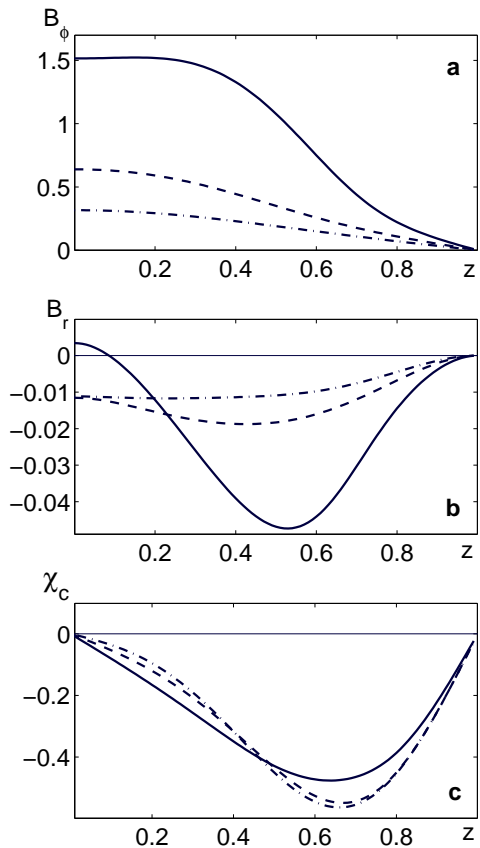}
\caption{\label{Fig5} The $z$-dependence of solutions for the
local model; $h/l=5$; $C=0.1$, $D=-40$ and $|\chi^v(z)|=\sin^2(\pi
z/2)$. The various curves indicate results with $\kappa=0.05$
(dashed-dotted), $\kappa=0.1$ (dashed) and $\kappa=0.3$ (solid).
a) the toroidal magnetic field $B_{\phi}(z)$, b) the poloidal
magnetic field $B_{r}(z)$, c) the function $\chi^c(z)$.}
\end{figure}

\begin{figure}
\centering
\includegraphics[width=8cm]{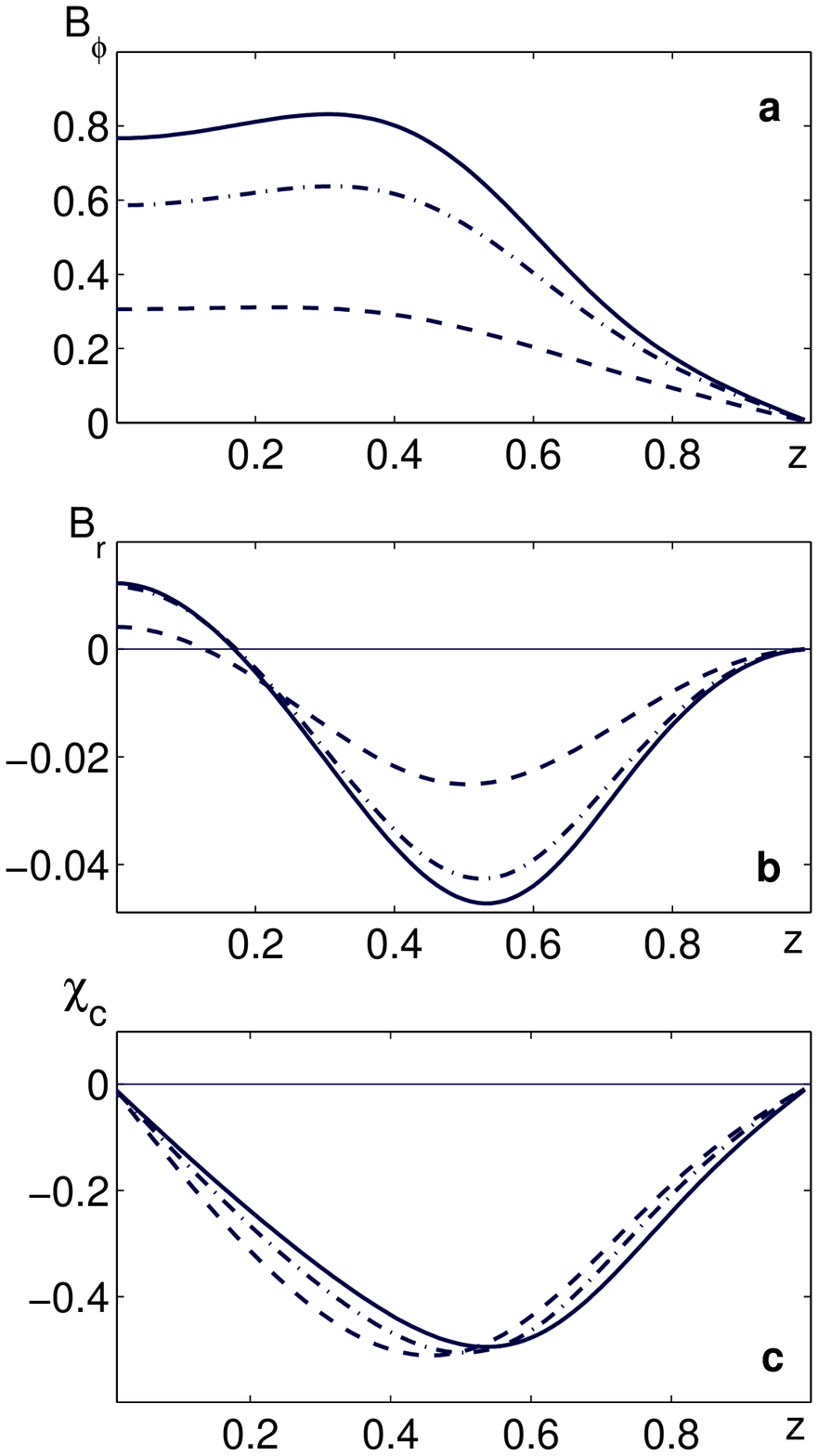}
\caption{\label{Fig6} The $z$-dependence of solutions for the
local model; $h/l=5$; $C=-0.1$, $D=-40$ and
$|\chi^v(z)|=\sin^2(\pi z/2)$. The various curves indicate results
without $\kappa$-quenching for $\kappa=0.1$ (dashed) and
$\kappa=0.3$ (solid) and with $\kappa$-quenching for $\kappa=0.3$
(dashed-dotted). a) the toroidal magnetic field $B_{\phi}(z)$, b)
the poloidal magnetic field $B_{r}(z)$, c) the function
$\chi^c(z)$.}
\end{figure}

New features were found in comparison with the results of Kleeorin
et al. (2002), where the quenching of the turbulent magnetic
diffusion was not taken into account. In particular, the maximum
of the toroidal magnetic field for $|D| > 40$ is not located at
$z=0$ but is shifted to $z \approx 0.5 h$ (see Figs.~1a and 3a).
The equipartition toroidal field is attained for smaller values of
the dynamo numbers and parameter $C$ than in the model studied by
Kleeorin et al (2002). The reason is that the asymptotic behavior
of the steady-state solution is different in these two cases: in
the present study $ B_\phi \propto  C \, |D| \, B_{\rm eq} $ (see
Eq.~(\ref{C6})) whereas when the quenching of the turbulent
magnetic diffusion vanishes we have $ B_\phi \propto  \sqrt{C \,
|D|} \, B_{\rm eq} $ (see Eq.~(18) in Kleeorin et al., 2002). Note
that for galaxies reasonable estimates are $|D| \sim 10-30$, $\,
|C| \sim 0.1$ and $\kappa \sim 0.3-0.5 .$ With these parameters
the present model gives toroidal field strengths of about the
equipartition value.

Fig.~2c demonstrates the change of sign of the effective drift
velocity with $z ,$ so that in one part of the disc it is
diamagnetic, and in the other it is paramagnetic. A diamagnetic
velocity implies that the field is pushed out from regions with
stronger mean magnetic field, while a paramagnetic velocity causes
the magnetic field to be concentrated in regions with stronger
field.

Fig.~6  shows the solutions for different values of the turbulent
diffusion $\kappa $ of the magnetic helicity. It is apparent from
Fig.~6 that the magnitude of the saturated toroidal magnetic field
increases with $\kappa .$ The reason is clear; the increase of
this parameter increases the flux of the magnetic helicity, and
causes a decrease of the magnetic part of the $\alpha$--effect,
thus increasing the total $\alpha$--effect.

When comparing the numerical  and asymptotic solutions we need to
take into account that the asymptotic solution
(\ref{C6})--(\ref{C7}) was obtained only for $\kappa=0.$ Thus such
a comparison can only be performed for very small values of
$\kappa .$ If we compare the field $B_\phi$ at $z=0$ for, e.g.,  $
C = 0.1$ and $ \kappa=0.05$ we find that the difference between
the asymptotic and numerical solutions is about 15 percent when $
D=-100 .$ The novel feature,  the maximum of the toroidal field
$B_\phi$ at $z>0$ rather than $z=0$, appears in the numerical
solutions for $\kappa \ge 0.1 $, and is not described in the
framework of the above asymptotic analysis. Note also that a
discrepancy between the numerical and asymptotic solutions is
perhaps not so surprising even for large values of $|D|$, as
relation~(\ref{C7}) diverges near $z=1 .$

In order to separate the study of the algebraic and dynamic
nonlinearities we define a function $ D_V(B) / D = \phi_v(B) /
[\eta_{A}(B) \eta_{B}(B)] $, using only the hydrodynamic part of
the $ \alpha $ effect. Thus, for  $ B \gg 1/ \sqrt{8} $ the
function $ D_V(B) / D $ is a nonzero constant, because then
$ \eta_{A}(B) \sim 2 / 5 \beta ,$ $ \eta_{B}(B)
\sim 3 / 5 \beta ,$ $ \phi_{v}(B) \sim 2 / \beta^{2} $, with
$\beta = \sqrt{8} B .$ The saturation of the growth of the mean
magnetic field in the case with only an algebraic nonlinearity
present can be achieved when the derivative of the function $ d
D_V(B) / dB < 0 .$ Thus, taking into account quenching of the
turbulent magnetic diffusion we find that the algebraic
nonlinearity alone (i.e. quenching of both the $ \alpha $ effect
and turbulent magnetic diffusion) cannot saturate the growth of
the mean magnetic field because $ d D_V(B) / dB \to 0 $ for $ B
\gg 1/ \sqrt{8} $.

We will show here that the combined effect of the algebraic and
dynamic nonlinearities limits the growth of the mean magnetic
field. The dynamic nonlinearity is determined by the evolutionary
equation~(\ref{A6}) for $ \chi^c $. We introduce the parameter $
D_N(B) / D = \alpha(B) / [\alpha(B=0) \eta_{A}(B) \eta_{B}(B)] $
which characterizes both the algebraic and dynamic nonlinearities,
while the parameter $ D_V(B) / D $ characterizes only the
algebraic nonlinearity. The saturation of the growth of the mean
magnetic field is achieved when the derivative of the nonlinear
dynamo number satisfies $ d D_N(B) / dB < 0 .$ We see in Fig.~7
that this condition is satisfied. However, we see also that $ d
D_V(B) / dB > 0 .$ This implies that the algebraic nonlinearity
alone (i.e. quenching of both the $ \alpha $ effect and turbulent
magnetic diffusion) cannot saturate the growth of the mean
magnetic field. The same follows from the above asymptotic
analysis.

\begin{figure}
\centering
\includegraphics[width=8cm]{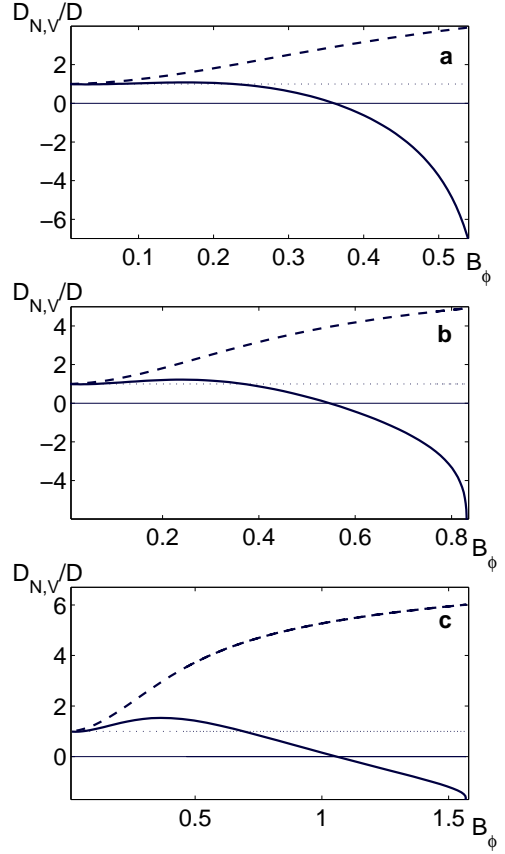}
\caption{\label{Fig7} The dependencies of $ D_N(B) / D = \alpha(B)
/ [\alpha(B=0) \eta_{A}(B) \eta_{B}(B)] $ (solid) and $ D_V(B) / D
= \phi_v(B) / [\eta_{A}(B) \eta_{B}(B)] $ (dashed) on the mean
magnetic field for the one-dimensional model; $h/l=5$; $C=-0.1$;
$\kappa=0.3$; with different $D$: a) $D=-20$; b) $D=-40$; c)
$D=-100$.}
\end{figure}

\section{Simple models with radial extent}

Detailed numerical modelling of  galactic dynamos is quite a
complicated numerical problem. Dynamo models with conventional
$\alpha$-quenching are however very robust and allow  drastic
simplifications that nevertheless  reproduce adequately the basic
features of galactic magnetic field structure as reflected in the
observational data. The aim of this section is to discuss to what
extent these simplified models are comparable with the quenching
discussed above.

The basic no-$z$ dynamo model for disc galaxies has proved to be a
useful tool for studying dynamo action in these objects, and is
described in Moss (1995). Here we also include the tuning
suggested by Phillips (2001), namely the multiplication by factors
$f=\pi^2/4$ of the terms representing the $z$-diffusion of $B_r$
and $B_\phi$.  The no-$z$ model differs from the local model of
Sect.~\ref{basic} in that it describes magnetic fields over the
entire radial range, $0\leq r\leq R$, but all explicit dependence
on the vertical coordinate $z$ has been removed, with the
first-order $z$-derivatives being replaced by $1/h$ and the
second-order $z$-derivatives being replaced by $- f/h^2$. The
field components $B_r$, $B_\phi$ appearing in the no-$z$ equations
can either be thought of as representing mid-plane values, or as
some sort of vertical average of values through the disc.

For the no-z model in the axisymmetric case the mean-field dynamo
equations take the form
\begin{eqnarray}
{\partial B_\phi \over \partial t} &=& - f \eta_B\biggl({B \over
\sqrt{\rho}}\biggr) B_\phi
\nonumber\\
& & + \lambda^2 {\partial \over \partial r} \biggl[
\eta_B\biggl({B \over \sqrt{\rho}}\biggr) {1 \over r} {\partial
\over \partial r} (r B_\phi) \biggr]
\nonumber\\
& & + R_\omega B_r  r {\partial \Omega \over \partial r} +
\lambda^2 {\partial \over \partial r} \biggl[ \eta_V \biggl({B
\over \sqrt{\rho}}\biggr) {B \over r} \biggr]\;,
\label{E12}\\
{\partial B_r \over \partial t} &=& - R_\alpha \alpha B_\phi -
\biggl[(f+1)\eta_A\biggl({B \over \sqrt{\rho}}\biggr)
\nonumber\\
& & - \, \eta_B\biggl({B \over \sqrt{\rho}}\biggr)\biggr] B_r +
\lambda^2 \eta_A\biggl({B \over \sqrt{\rho}}\biggr) {\partial
\over \partial r} \biggl({1 \over r} {\partial \over \partial r}
(r B_r) \biggr)
\nonumber\\
& & - \lambda^2 \tilde V_A\biggl({B \over \sqrt{\rho}}\biggr) {1
\over r} {\partial \over \partial r} (r B_r) \; .
\label{E14}\\
{\partial \chi^c \over \partial t} &=& {1 \over \rho(r)} \biggl \{
4 \biggl({h \over l} \biggr)^2 \biggl[C \, \chi^v \,
\eta_A\biggl({B \over \sqrt{\rho}}\biggr) \, \phi_v\biggl({B \over
\sqrt{\rho}}\biggr) \, B^2
\nonumber\\
&& - \alpha B^2 \biggr] - \kappa  \chi^c + {4 \lambda^2 \over r}
{\partial \over \partial r} \biggl(\kappa \, r \, {\partial \chi^c
\over
\partial r} \biggr)
\nonumber\\
&& + {4\lambda^2  \over r} {\partial \over
\partial r} \biggl[ r \, \Lambda_{\rho}^{-1} \, \chi^v \,
\eta_A\biggl({B \over \sqrt{\rho}}\biggr) \, \phi_v\biggl({B \over
\sqrt{\rho}}\biggr) \, B^2 \biggr]
\nonumber\\
&& + 4 \lambda^2 \, \eta_A\biggl({B \over \sqrt{\rho}}\biggr) \, S
\biggr\} , \label{E15}
\end{eqnarray}
where
\begin{eqnarray*}
\tilde V_A &=& [\eta_A - \eta_B] {\partial \over \partial r} \ln
\biggl({B \over \sqrt{\rho}}\biggr) + {1 \over 2r} A_2(\beta) \;,
\\
S &=& - {1 \over r^2} {\partial \over \partial r} (r B_r) \,
{\partial \over \partial r} (r B_\phi) + B_\phi {\partial \over
\partial r} \biggl({1 \over r} {\partial \over \partial r} (r B_r)
\biggr) \;,
\end{eqnarray*}
and $\, \lambda= h/R$ is the aspect ratio, $R_\alpha=\lambda \,
\alpha_0 \, h/\eta_0 ,$ $\, \Lambda^{-1}_\rho = |\vec{\nabla}_r \,
\rho| / \rho $ and $\eta_0$ is the maximum (unquenched) value of
$\eta$. The quenching functions $\phi_v$, $\phi_{m}$, $\eta_A$,
$\eta_B$ and $\eta_V$ contain in their arguments the factor
$1/\sqrt{\rho}$ because they are based on local equipartition at
radius $r$. To determine the magnetic field distribution along the
radius we use a Brandt rotation law, $\Omega(r) = \Omega_0 /[1 +
(r/r_\omega)^2]^{1/2}$ with $r_\omega = 0.2 ,$ and the radial
density profile $\rho(r) = \exp[-(r/r_\rho)^2]$ with $r_\rho =
0.5$, so that $\Lambda_\rho = 2r/r_\rho^2 .$ We also set
$\chi^v(r) =1.$

When comparing the no-$z$ model with the local thin-disc model
studied in the earlier parts of this paper, note that in the local
thin-disc model, $R_\omega=r h^2 \eta_T^{-1} d\Omega/dr<0$.  By
the nature of the model, $rd\Omega/dr$ is the value at a chosen
radius in the disc, and $r$ does not further occur explicitly in
the analysis. However the no-$z$ model is {\it global} with
respect to radius, and the value of $rd\Omega /dr$ varies through
the disc, from zero at $r=0$ to some maximum absolute value; for
the Brandt rotation law this value is $0.31\Omega_0$ at
$r=\sqrt{2/3}r_\omega$. For the no-$z$ model the global definition
is $R_\omega=R_\omega({\rm no}-z) = \Omega_0 h^2 / \eta_T>0 $.
Less importantly, there are also small differences, of order 25\%,
in the effective values of $R_\alpha$ occurring in the two
approximations, even though the formal definitions are the same --
see Phillips (2001).

We investigated solutions to these no-$z$ equations for a range of
parameter values. However we were unable to obtain satisfactory
convergence to finite solutions without including a contribution
to the diffusivity that was not quenched. Our feeling is that the
simplest form of the no-$z$ formalism may not be sufficiently
robust to allow inclusion of the full $\eta$-quenching formalism
described above. This is in sharp contrast to the situation
without the inclusion of $\eta$-quenching (Kleeorin et al. 2002),
when satisfactory solutions with approximately equipartition
strength fields were readily obtained.

However, given the evidently satisfactory behavior of the local
model presented in Sect.~4, it is apparent that extension to the
radial range $0\le r\le R$ in the manner described by Ruzmaikin et
al. (1988, Ch.~7) would present no difficulties in principle, nor
new features.

\section{Discussion and conclusions}

In this paper we present a more detailed description of a
nonlinear galactic dynamo, that includes quenching of the turbulent
diffusivity of the magnetic field in addition to the effects
considered in our previous paper (Kleeorin et al. 2002).
We find that as far as the model leads to results that are
comparable with observations, these results are similar to those
obtained from conventional galactic dynamo models, with
large-scale magnetic fields typically of equipartition strength,
and with plausible values of the pitch angles.
We confirm the conclusion of Kleeorin et al. (2002) that from a
pragmatic point of view conventional models of nonlinear galactic
dynamos are quite adequate to reproduce the directly observable
manifestations of galactic dynamo action.

Our approach is based on first principles as far as is possible
in the framework of mean-field dynamo theory, and results in the
conclusion that the self-consistent form of dynamo saturation is
much more complicated than is suggested in conventional
models for a  galactic dynamo. We demonstrate the important role of
two types of nonlinearity (algebraic and dynamic) in the
mean-field galactic dynamo. The algebraic nonlinearity is
determined by a nonlinear dependence of the mean electromotive
force on the mean magnetic field. The dynamic nonlinearity is
determined by a differential equation for the magnetic part of the
$ \alpha $-effect. This equation is a consequence of the
conservation of the total magnetic helicity. We have taken into
account the algebraic quenching of both the $ \alpha $ effect and
the turbulent magnetic diffusion, and also dynamic nonlinearities
(see e.g. Rogachevskii \& Kleeorin 1999, 2000, 2001). Since the
quenching of the $ \alpha $ effect and the turbulent magnetic
diffusion have the same origin (i.e. the direct and indirect
modification of the electromotive force by the mean magnetic
field), they cannot in general be taken into account separately.
This implies there is no reason to include $ \alpha $ quenching
and to ignore the quenching of the turbulent magnetic diffusion,
or {\it vice versa}.

We have also verified that the algebraic nonlinearity alone (i.e.
quenching of both the $ \alpha $ effect and turbulent magnetic
diffusion) cannot saturate the growth of the mean magnetic field.
The situation changes when the dynamic nonlinearity is taken into
account. The crucial point is that the dynamic equation for the
magnetic part of the $ \alpha $-effect (i.e. the dynamic
nonlinearity) includes the flux of magnetic helicity. Without this
flux, the total magnetic helicity is conserved locally and the
strength of the saturated mean magnetic field is very small
compared to the equipartition strength. The inclusion of a
magnetic helicity flux means that the total magnetic helicity is
not conserved locally because the magnetic helicity of small-scale
magnetic fluctuations is redistributed by the helicity flux. In
this case an integral of the total magnetic helicity over the disc
is conserved. The equilibrium state is given by a balance between
magnetic helicity production and magnetic helicity transport (see
Kleeorin et al. 2002). Thus, the combined effect of algebraic and
dynamic nonlinearities limit the growth of the mean magnetic field
and results in an equilibrium strength of the mean magnetic field
which is of order that of the equipartition field, in agreement
with observations of galactic magnetic fields.

We find, perhaps quite naturally, that when including new
physically significant effects we obtain less robust models. The
limitations of simulations with such models are connected not only
with purely numerical problems, which are however more severe than
for conventional galactic dynamo simulations. In order to
reproduce the detailed evolution of helicity and turbulent
diffusivity there is the necessity for a deeper description of the
real multiphase structure of  the interstellar medium, and the
physical processes that result in the development of helical
interstellar turbulence, than the standard description used in
present-day models. Of course, the corresponding development of
mean-field dynamo theory required to achieve this goal is far
beyond the scope of this paper. We stress that direct numerical
simulations of interstellar turbulence (see e.g. Korpi et al.
1999), followed by estimations of the appropriate mean-field
dynamo control parameters, are very desirable and even essential
in this context.

\begin{acknowledgements}
Financial support from NATO under grant PST.CLG 974737, RFBR under
grant 01-02-16158 and the INTAS Program Foundation under grant
99-348 is acknowledged. DS is grateful to a special fund for
visiting senior scientists of the Faculty of Engineering of the
Ben-Gurion University of the Negev, and to the Royal Society, London,
for financial support during his visits to Israel and the UK respectively.
\end{acknowledgements}


\begin{thebibliography}{}

\bibitem[1996]{}
Beck, R., Brandenburg, A., Moss, D., Shukurov, A.,  \& Sokoloff,
D. 1996, ARA\&A, 34, 155

\bibitem[1999]{}
Beck R., Ehle M., Shoutenkov V., Shukurov A., \& Sokoloff D. 1999,
Nature, 397, 324

\bibitem[2002]{}
Beck R., Shoutenkov V., Ehle M., Harnett J. I., Haynes R. F.,
Shukurov A., Sokoloff D., \& Thierbach M. 2002, A\&A, 391, 361

\bibitem[2002]{}
Blackman, E. G., \& Brandenburg, A. 2002, ApJ. 579, (scheduled for
the November 1 issue)

\bibitem[2001]{}
Brandenburg A., 2001, ApJ 550,824

\bibitem[2001a]{}
Brandenburg, A., Bigazzi, A.,  \& Subramanian, K. 2001a, MNRAS,
325, 685

\bibitem[2001]{}
Brandenburg A., \& Dobler W. 2001, A\&A, 369, 329

\bibitem[2002]{}
Brandenburg, A., \& Sokoloff, D. 2002, GAFD, 96, 319

\bibitem[1999]{}
Brandenburg A., \& Subramanian K. 2000, A\&A, 361, L33

\bibitem[2001]{}
Chae, J. 2001, ApJ, 560, L95

\bibitem[1998]{}
Covas, E., Tavakol, R., Tworkowski, A.,  \& Brandenburg, A. 1998,
A\&A, 329, 350

\bibitem[2002]{}
Dobler, W., Shukurov, A., \& Brandenburg, A. 2002, Phys. Rev. E
65, 036311

\bibitem[1999]{}
Field, G., Blackman, E.,  \& Chou, H. 1999, ApJ, 513, 638


\bibitem[1975]{}
Frisch U., Pouquet A., Leorat I.,  \& Mazure A., 1975, J. Fluid
Mech. 68, 769

\bibitem[1995]{}
Gruzinov, A. V.,  \& Diamond, P. H. 1995, Phys. Plasmas, 2, 1941

\bibitem[1970]{}
Iroshnikov, R. S. 1970, SvA, 14, 582; 1001

\bibitem[1999]{}
Kleeorin, N.,  \& Rogachevskii, I. 1999, Phys. Rev. E., 59, 6724

\bibitem[2000]{}
Kleeorin, N., Moss, D., Rogachevskii, I.,  \& Sokoloff, D. 2000,
A\&A,  361, L5

\bibitem[2002]
Kleeorin, N., Moss, D., Rogachevskii, I.,  \& Sokoloff, D. 2002, A\&A,
387, 453

\bibitem[1995]{}
Kleeorin, N., Rogachevskii, I.,  \& Ruzmaikin, A. 1995, A\&A, 297,
159

\bibitem[1982]{}
Kleeorin, N.,  \& Ruzmaikin, A., 1982, Magnetohydrodyn.,  18, 116

\bibitem[1999]{}
Korpi M., Brandenburg A., Shukurov A., Tuominen I., \& Nordlund
\AA., 1999, ApJ 514, L99

\bibitem[1980]{}
Krause F., \& R\"adler K.-H. 1980,  Mean-Field
Magnetohydrodynamics (Oxford, Pergamon)

\bibitem[1999]{}
Kulsrud, R. 1999, Ann. Rev. Astron. Astrophys., 37, 37

\bibitem[1995]{}
Moss, D. 1995, MNRAS 275, 191

\bibitem[2001]{}
Moss D., Shukurov A., Sokoloff D., Beck R., \& Fletcher A., 2001,
A\&A 380, 55

\bibitem[2001]{}
Phillips, A. D., 2001, GAFD 94, 135

\bibitem[2000]{}
Rogachevskii, I., \&  Kleeorin, N. 2000,  Phys. Rev. E., 61, 5202

\bibitem[2001]{}
Rogachevskii, I., \& Kleeorin, N. 2001,  Phys. Rev. E., 64, 056307

\bibitem[1988]{}
Ruzmaikin, A., Shukurov, A., \& Sokoloff, D. 1988, Magnetic Fields
of Galaxies (Kluwer, Dordrecht)

\bibitem[1991]{}
Schmalz, S., \& Stix, M. 1991, A\&A, 245, 654

\bibitem[1992]{}
Vainshtein, S. I., \& Cattaneo, F. 1992, ApJ,  393, 165

\bibitem[2001]{}
Vishniac, E. T., \& Cho, J. 2001, ApJ,  550, 752

\bibitem[1983]{}
Zeldovich, Ya. B., Ruzmaikin, A. A., \& Sokoloff, D. D. 1983,
Magnetic Fields in Astrophysics (Gordon and Breach, New York)

\end{thebibliography}
\end{document}